\documentclass[prb,twocolumn,showpacs]{revtex4}

\usepackage{graphicx}
\usepackage{dcolumn}
\usepackage{bm}

\begin{document}

\draft
\title{Nonlinear sliding friction of adsorbed overlayers on
disordered substrates}
\author{Enzo Granato}
\address{Laborat\'orio Associado de Sensores e Materiais,
Instituto Nacional de Pesquisas Espaciais, \\
12245-970 S\~{a}o Jos\'e dos Campos, S\~ao Paulo, Brazil}
\author{S.C. Ying}
\address{Department of Physics,
Brown University, \\
Providence, Rhode Island 02912}

\begin{abstract}
We study the response of an adsorbed monolayer on a disordered
substrate under a driving force using Brownian molecular-dynamics
simulation. We find that the sharp longitudinal and transverse
depinning transitions with hysteresis still persist in the
presence of weak disorder. However, the transitions are smeared
out in the strong disorder limit. The theoretical results here
provide a natural explanation for the recent data for the
depinning transition of Kr films on gold substrate.

\draft

\end{abstract}

\pacs{68.35.Gy, 68.35.Rh, 81.40.Pq}

\maketitle

\section{Introduction}

Adsorbed overlayers under a driving force have attracted much
interest recently
\cite{proc,book,persson,robbins,braun,gy,gy00,das,gao,vanossi,diestler}
as important examples of driven lattice systems
\cite{gled,balents}. The nonlinear response of the overlayer in
the presence of pinning centers resulted in various steady states,
corresponding to a variety of different nonequilibrium dynamic
phases. The nature of these dynamic phases and the transitions
from one phase into each other are the microscopic factors
controlling the macroscopic sliding friction and boundary
lubrication \cite{book,exp,exp2,carlin02,carlin}. Other closely
related systems include vortex lattices in type-II superconductors
\cite{moon,vl,nori,gty,olson,carneiro}, charge-density waves
\cite{gruner} and colloidal crystal layers \cite{ling}. In
general, both a periodic potential and disorder due to quenched
defects are present in theses systems. In particular, in the
sliding friction and boundary lubrication of surfaces the
underlying crystal lattice provides the periodic pinning potential
which competes with defect pinning centers due to impurities.

For periodic substrates, it has been shown through molecular
dynamics simulation studies  that there are different dynamical
phases for the driven overlayer such as pinned solid, sliding
solid and liquid phases. The transitions between these phases
often involve strong hysteresis effects \cite{persson}, leading to
important macroscopic behavior as the stick and slip motion, which
can be observed experimentally with the surface force apparatus
\cite{exp,exp2}. Most of the theoretical results were based on
two-dimensional models of overlayers with Lennard-Jones potential
interaction between the adatoms.\cite{persson}. The results on
these dynamical phases have been found to be universal and are
reproduced even in models with pure elastic interacting potential
\cite{gy}. More recently \cite{gy00}, it has also been shown that,
in the sliding state, the response of the overlayer to an
additional force in the transverse direction can also lead to a
depinning transition at low temperatures. Different sliding states
also occur in presence of quasiperiodic substrates and have been
investigated in detail recently for one-dimensional models
\cite{gao,vanossi}. However, the effects of disorder on the main
features observed for two-dimensional systems, which can arise,
for example, from point defects on a initially periodic substrate,
have not been investigated satisfactorily so far. Considerable
analytical and numerical results are only available for driven
lattice systems described by overdamped dynamics
\cite{gled,balents}. Unfortunately, it is known that inertial
effects  significantly modify the nature of the depinning
transition \cite{book}. Hence the results for these overdamped
systems cannot be directly applied to the sliding friction problem
involving adsorbed overlyers on a substrate.

In this work, we study numerically the nonlinear sliding friction
of adsorbed overlayers on disordered substrates with a specific
model of point defects, where the strength of the disorder can be
varied by changing the concentration of the defects. We
investigate the longitudinal and transverse response for weak and
strong disorder by Brownian molecular dynamics simulations. We
find that the main features, such as hysteresis, longitudinal and
transverse depinning transitions previously observed for periodic
surfaces \cite{persson,gy00}, still survive in the presence of
weak disorder but are smeared out in the strong disorder limit.
Recently, the depinning transition of  Kr films on gold has been
studied using the quartz crystal microbalance technique
\cite{carlin}. Our results for the longitudinal response bear a
strong qualitative resemblance to the experimental data of the
depinning transition. In particular, it provides concrete
theoretical  support to the speculation
 that the aging effect on the depinning transition is
due to the increasing disorder of the substrate.

\section{Model and simulation}

The model of interacting adatoms on a substrate with point
disorder subject to an external driving force for the present
study   is a simple extension of the one previously used for
periodic substrates \cite{book,persson,gy00}. The dynamics is
described by the Langevin equation
\begin{equation}
m\ddot  {\bf r}_i + m \eta \dot{\bf r}_ i = - \frac{\partial
U}{\partial{\bf r}_i} - \frac{\partial V}{\partial{\bf r}_i} +
{\bf f}_i + {\bf F}
\end{equation}
where ${\bf r_i=(x_i,y_i})$ is the adsorbate position, $U=\sum_i
u(r_i)$ is the  substrate pinning potential, $V=\sum_{i\ne j}
v(|{\bf r_i}-{\bf r_j})$ is the interactions between adatoms,
${\bf F}$ is the uniform external force acting on each adatom and
${\bf f}_i$ is a stochastic force, with zero average, and variance
related to the microscopic friction parameter $\eta$, the mass of
the particles $m$ and the temperature $T$ by the fluctuation
dissipation relation
\begin{equation}
<f_i^{\alpha}(t)f_i^{\alpha}(t')  > = 2 \eta m kT \delta(t-t') ,
\end{equation}
where $\alpha$ represents Cartesian components. We choose a
substrate potential  which provides a periodic pinning for the
overlayer except near a pinning center, given by the form
\begin{equation}
u({\bf r}) = U_0 \ [2-\cos(2\pi x/a) - \cos(2 \pi y/a)]
\label{ppot}
\end{equation}
\label{periodic}
Near a pinning center (point defect) located at
$R_p = (X_p,Y_p)$, the substrate potential is locally modified to
the form
\begin{equation}
u({\bf r}) = U_p \ [2-\cos(2\pi x/a) - \cos(2 \pi y/a)]
\end{equation}
for
\begin{eqnarray}
 && X_p-a/2 < x < X_p +a/2  \cr
 && Y_p-a/2 < y < Y_p +a/2 ,
\end{eqnarray}

The strength of the random pinning potential $U_p$ is chosen to be
uniformly  distributed in a range of $4U_0$ with an average
$<U_p>=U_o$. The position of the pinning centers $R_p = (X_p,Y_p)$
is randomly chosen on a square lattice of lattice parameter $a$
and linear size $L$. With this choice, a pinning center consists
of a square region with a force that vanishes smoothly at the
boundary. For $U_p=U_o$, a constant, the entire substrate
potential reduces to the periodic cosine potential in Eq.
\ref{ppot}  studied previously \cite{book,persson,gy00}. The
strength of disorder is controlled  by varying the concentration
$x_d$ of randomly chosen pinning centers. In the dilute limit of $
x_d << 1$, corresponding to weak disorder, this pinning potential
has a predominant Fourier component at the wavevector $\vec k=(
2m\pi/a, 2n\pi/a) $. In the other limit of dense pinning centers
with $ x_d \rightarrow 1$, it represents an amorphous substrate
with strong disorder. The interaction between adatoms is
represented  by a Lennard-Jones pair potential
\begin{equation}
v(r_{ij}) = \epsilon [(r_o/r_{ij})^{12} - 2 (r_o/r_{ij})^6 ]
\end{equation}
where $\epsilon$ is the well depth and $r_0$ is the particle
separation at the minima in the pair potential. We have chosen the
parameters in the interaction potential to be $r_o/a \sim 1.5$,
$\epsilon=U_0$, and the adsorbate coverage  at $\theta=1/2$. This
choice is made such that the ground state of  the overlayer in the
periodic substrate ($x_d=0$ ) corresponds to the commensurate
pinned $c(2 \times 2)$ structure studied previously. For the
periodic substrate, the longitudinal and transverse response of
the commensurate overlayer to an applied force in the present
model have been extensively studied in connection with the sliding
friction and boundary lubrication \cite{book,persson,gy00}
problem. Thus, this model provides a convenient starting point for
the present study of disorder effects.  In the discussions below
we use dimensionless units   normalizing the force by $2\pi U_o$,
velocities by $2\pi U_o/\eta$ and temperature by $U_0/k_B$. We
study the nonlinear response of the overlayer at different
temperatures and pinning center concentrations via Brownian
molecular dynamics simulations \cite{allen}. Systems consisting of
$N = 50 $ to $ 800$ adatoms were considered with the time variable
discretized in units of $\delta t=0.002 - 0.01 \tau$, where
$\tau=(ma^2/U_o)^{1/2}$. The main results were obtained for
$\epsilon = U_o$ and $\eta=1$. For each set of parameters,
calculations were performed for typically $ 1-4 \times 10^5$ time
steps to allow the system to reach steady state. Then an equal
number of time steps are used to evaluate time-averaged values of
various physical quantities in this steady state. Finally, the
results are also averaged over a number of (typically 6-11)
different realizations of the disorder through the distribution of
the $U_p$ value for the pinning centers.

\section{Results and Discussion}
\subsection{Depinning Transition and Longitudinal Response}
\begin{figure}
\includegraphics[bb=1cm 3cm 20cm 28cm, width=8cm]{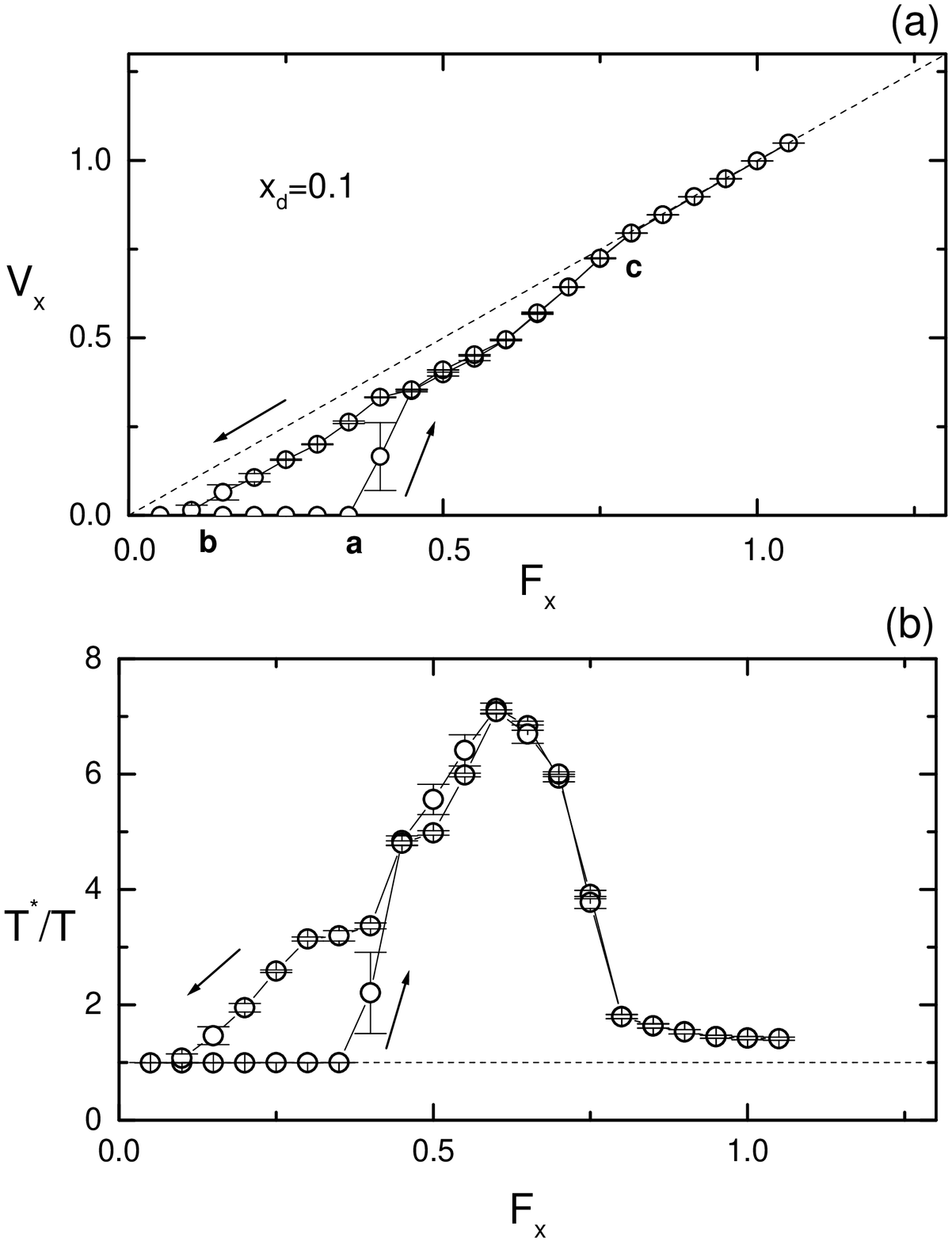}
\caption{Drift velocity $V_x$ (a) and effective temperature $T^*$
(b) of the overlayer as a function of the external force $F_x$,
for dilute disorder $x_d=0.1$. The results are for $T=0.2$ and
system size $L=20$. The direction of variation of $F_x$ is
indicated by arrows. } \label{lx01}
\end{figure}

\begin{figure}
\includegraphics[bb=1cm 3cm 20cm 28cm, width=8cm]{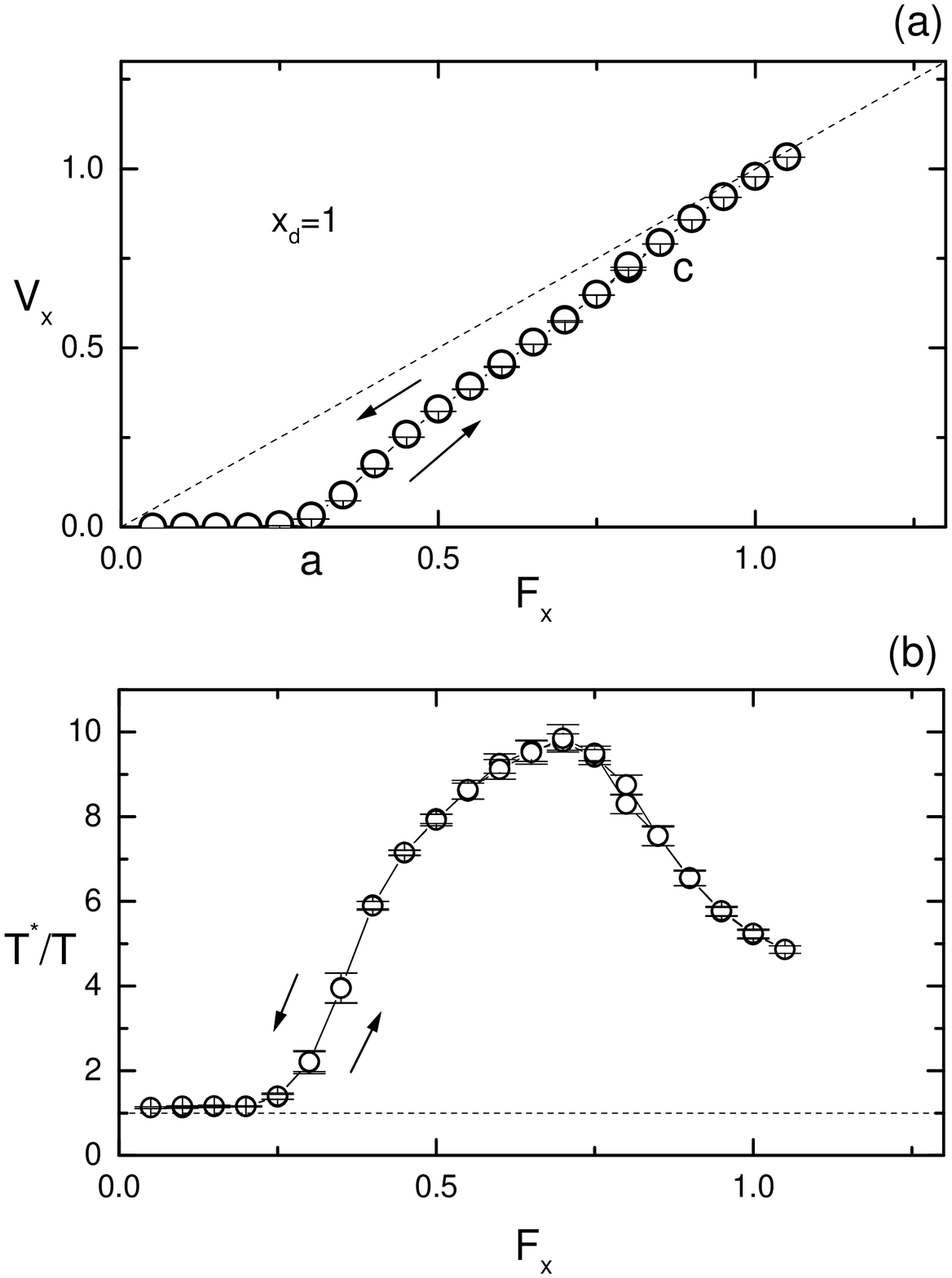}
\caption{Drift velocity $V_x$ (a) and effective temperature $T^*$
(b) of the overlayer  as a function of the external force $F_x$,
for dense disorder $x_d=1$. The results are for $T=0.2$ and system
size $L=20$. The direction of variation of $F_x$ is indicated by
arrows. } \label{lx1}
\end{figure}

The longitudinal response to an applied force at a temperature
$T=0.2$, well below the melting transition temperature of the $c(2
\times 2)$ phase on a periodic substrate, $T_m=1.2$, is shown in
Figs. \ref{lx01} and \ref{lx1} for dilute $x_d=0.1$ and dense
disorder $x_d=1$, respectively. In the former case, the main
features are similar to that found for the periodic substrate
\cite{persson,gy00}. The drift velocity $V_x$ is essentially zero
below a critical value of $F_a$ (neglecting the tiny contribution
from thermal creep motion which is always present for nonzero
force). When the driving force exceeds the critical value $F_a$,
there is a sharp depinning transition where the overlayer starts
to slide and the drift velocity increases rapidly. Eventually
 a sliding regime for $F_x >> F_c$ is reached where
the longitudinal sliding friction, defined as $\bar \eta =
F_x/V_x$, is close to the microscopic friction, $\bar \eta \sim
\eta $. There exists a hysteresis behavior for the depinning
transition with unequal critical forces $F_a$ and $F_b$ for
increasing and decreasing external force $F_x$, corresponding to
the static and kinetic friction forces, respectively. On the other
hand, in the dense disorder limit $x_d=1$ of Fig. \ref{lx1}, the
hysteresis is completely washed out leading to equal static and
kinetic friction forces $F_a=F_b$.

\begin{figure}
\includegraphics[bb=1cm 3cm 20cm 28cm, width=8cm]{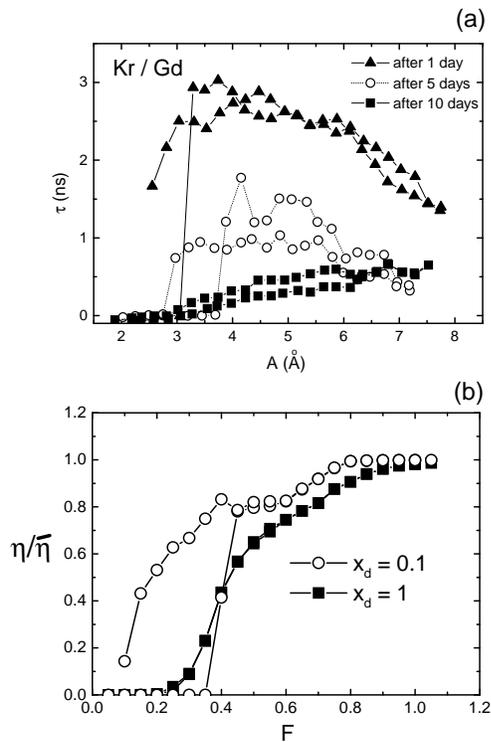}
\caption{ (a) Slip time $\tau$ as a function of the amplitude of
substrate oscillations, obtained from experiments on Kr/Gd films
using  the quartz crystal microbalance technique, by Carlin {\it
et al. } \cite{carlin}. (b) Inverse of the nonlinear sliding
friction $\bar \eta$ (longitudinal), normalized by the microscopic
friction parameter $\eta$,  as a function of the applied force for
weak ($x_d=0.1$) and strongly disordered ($x_d=1$) substrate,
obtained from the numerical simulations of this work. }
\label{hyst}
\end{figure}

Recently Carlin et al \cite{carlin} have studied the depinning
transition under an external driving force for Kr films on gold.
They employed the quartz crystal microbalance technique. In this
technique, the quartz crystal is driven to resonance by an
external rf voltage with the Kr film adsorbed on the gold
electrodes. The driving force on the admolecules  is proportional
to the  oscillation amplitude $A$ of the electrodes and the
resultant response of the Kr film is measured through the
"slipping time" $\tau$ which is inversely proportional to the
sliding friction $\bar \eta$. For a given coverage of the Kr film,
they found a sharp depinning transition when the oscillation
amplitude exceeds a critical value. This transition shows the same
kind of hysteresis effects as obtained in our weak disorder result
shown in Fig.\ref{lx01}. A very interesting result they found is
that the depinning transition changes as a function of aging of
the system. As shown in Fig. \ref{hyst}a, the depinning transition
is sharp and shows hysteresis behavior after 1 day. However, the
corresponding results after 10 days show a rounded transition
without hysteresis. The authors speculated that this change is
probably due to the deposition of foreign molecules onto the gold
electrode substrate, acting then as defect pinning centers. In
other words, the disorder of the substrate increases with aging.
In Fig. \ref{hyst}b, we combine the numerical results for
$x_d=0.1$ and $x_d=1$ from the present model to show the effect of
increasing density of defect pinning centers on the depinning
transition. Our results bear a strong qualitative resemblance to
the experimental data shown in Fig. \ref{hyst}a. In particular,
while the weak disorder ($x_d=0.1$) substrate shows a sharp
depinning transition with strong hysteresis effect, the depinning
transition for the strong disorder ($x_d=1$) substrate is smeared
out and there is no trace of hysteresis in this case. In fact, our
result for the strong disorder case bears a strong qualitative
resemblance to the experimental data aged after 10 days.  Thus,
the theoretical  results obtained in our present model study
clearly support the speculation \cite{carlin} that the aging
effect on the depinning transition observed in the quartz
microbalance experiment is due to the increasing disorder of the
substrate. Another interesting feature of the experimental results
in Fig. 3a is the monotonic decrease of the slip time with aging.
It should be noted that this effect is not seen in Fig. 3b because
the nonlinear sliding friction $\bar \eta $ is normalized by the
unknown microscopic friction parameter $\eta $, contained in the
effective model of Eq. (1). This parameter is essentially the
frictional damping of individual adatoms due to coupling to the
substrate. As the morphology of the substrate changes, the
frictional damping may change significantly. So the monotonic
decrease of the slip time in Fig. 3a can be accommodated within
our model as resulting from an increase in this microscopic
frictional damping with aging. The calculation of this microscopic
parameter will certainly require a detailed theoretical model for
this mechanism including realistic potentials which are outside
the scope of the present work.

\subsection{Nature of the sliding state}
\begin{figure}
\includegraphics[bb=1cm 3cm 20cm 28cm, width=8cm]{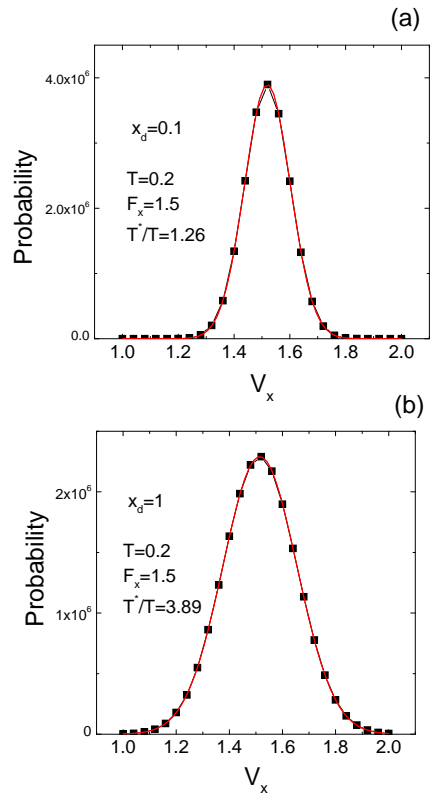}
\caption{ Velocity distribution at $T=0.2$, $F_x=1.5 $, for
$x_d=0.1$ (a)  and $x_d=1$ (b). Continuous line is a Gaussian
fit.} \label{vel}
\end{figure}

To understand the nature of the sliding state at different sliding
velocities, it is useful to introduce, just as in the periodic
case \cite{persson}, the concept of an effective temperature $T^*$
of the sliding overlayer. This is  defined via the velocity
fluctuation as
\begin{equation}
T^* = m (<{\bf v}^2>-<{\bf v}>^2)/2.       \label{effT}
\end{equation}
For a meaningful definition of this effective temperature, one
would expect that the velocity distribution  should be Gaussian
and the width of the distribution related to the effective
temperature by Eq. \ref{effT}. This is confirmed in Fig. \ref{vel}
where it is shown that the velocity distribution at $F_x
> F_c$ is in fact well described by a Gaussian both for
$x_d=0.1$ and $x_d=1$.

The behavior of the effective temperature $T^*$ under the driving
force carries important information for the structure properties
of the sliding state. As can be seen from Fig. \ref{lx01}, in the
dilute disorder case with $x_d=0.1$, this effective temperature is
much higher than $T$ in the region $ F_b < F_x < F_c$ and
decreases for $F_x >> F_c$, although never approaching $T$ as in
the periodic case. This suggests that for weak disorder, the
region for $ F_b < F_x < F_c$ is strongly disordered and
corresponds to a dynamically melted  liquid phase. For $F_x >>
F_c$, the sliding state is a solid state with weak disorder. For
the substrate with dense disorder $x_d=1$, the effective
temperature remains much higher than $T$ even for large driving
force, suggesting that in this case the overlayer remains in a
liquid like or floating state  even at large driving force and
high drift velocities. It is interesting to point out that an
expression relating $T^*-T$ and $\bar \eta - \eta$ proposed by
Persson \cite{persson} has been found to work well for a periodic
pinning potential. According to this relation, in the limit of
large driving force $F_x> F_c$, $\bar \eta$ should approach $\eta$
and $T^*$ should approach $T$. Our results in Figs. \ref{lx01} and
\ref{lx1} show that this expression breaks down in the presence of
disorder. While  $\bar \eta$ does approach $\eta$ in the large
driving force limit, $T^*$ remains higher than $T$ in this limit
in presence of disorder.

\begin{figure}
\includegraphics[bb=3cm 4cm 17cm 27cm, width=8cm]{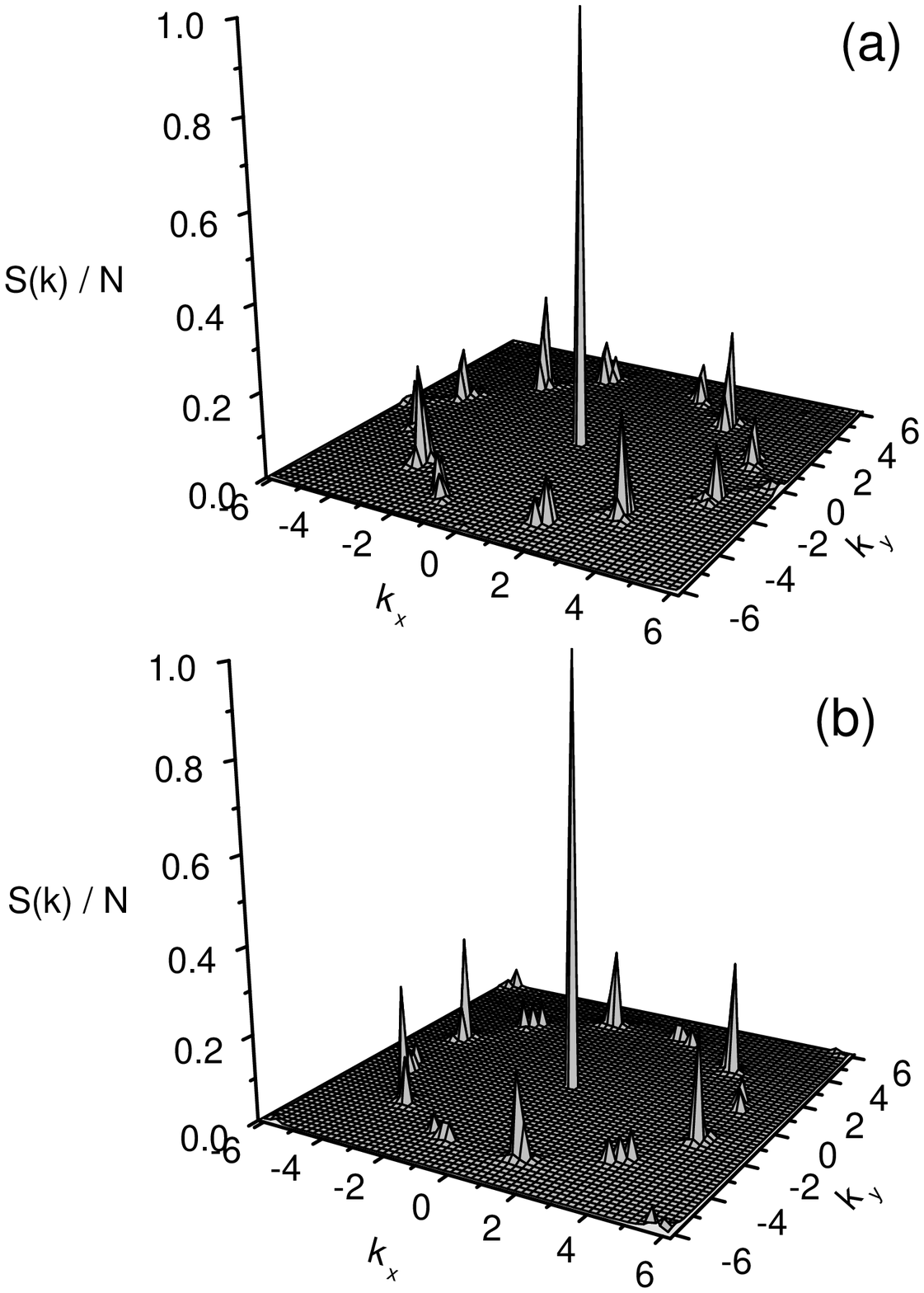}
\caption{ Structure factor $S(k)/N$ at $F_x=1.5 $, $F_y=0$ for
dilute disorder $x_d=0.1$ (a)  and dense disorder $x_d=1$ (b). The
results are for $T=0.2$ and system size $L=30$. } \label{sf}
\end{figure}

To support the above interpretation between the effective
temperature $T^*$ of the sliding overlayer and its structure, we
have evaluated  the structure factor defined by
\begin{equation}
S({\bf k}) = \frac{1}{N} [<\sum_{m,n} e^{i {\bf k} \cdot ({\bf
r}_m-{\bf r}_n )}>]
\end{equation}
for large driving force at $F_x=1.5$, $T=0.2$ in both the dilute
and dense disorder limit in the present model. The angular
brackets denote thermal averages and the square brackets disorder
averages. The results are shown in Fig. \ref{sf}. For a fully
ordered configuration the normalized structure factor $S(k)/N=1$
at the peak for a wavevector corresponding to the ordered
structure. As expected, for both $x_d=0.1$ and $x_d=1$ cases, this
value is much smaller at the nonzero wavevectors confirming that
the sliding layer in the presence of the disorder does not have
full long range order. For $x_d=0.1$ the structure factor shows
clear prominent peaks at wavevectors  corresponding to a
commensurate $c(2\times 2)$ structure, ($\vec k=(1,1)\pi/a$),
consistent with a weak disordered sliding solid state. For
$x_d=1$, corresponding to the strong disorder limit, one can
identify in the structure factor still well defined, although
weaker, peaks at the positions corresponding to an incommensurate
hexagonal structure. However, the information about the initial
periodic substrate no longer exists. The structure of the
overlayer in this strong disorder limit can be characterized as an
incommensurate floating state both in $x$ and $y$ directions
rather than as a liquid state.

\begin{figure}
\includegraphics[bb=3cm 4cm 17cm 27cm, width=8cm]{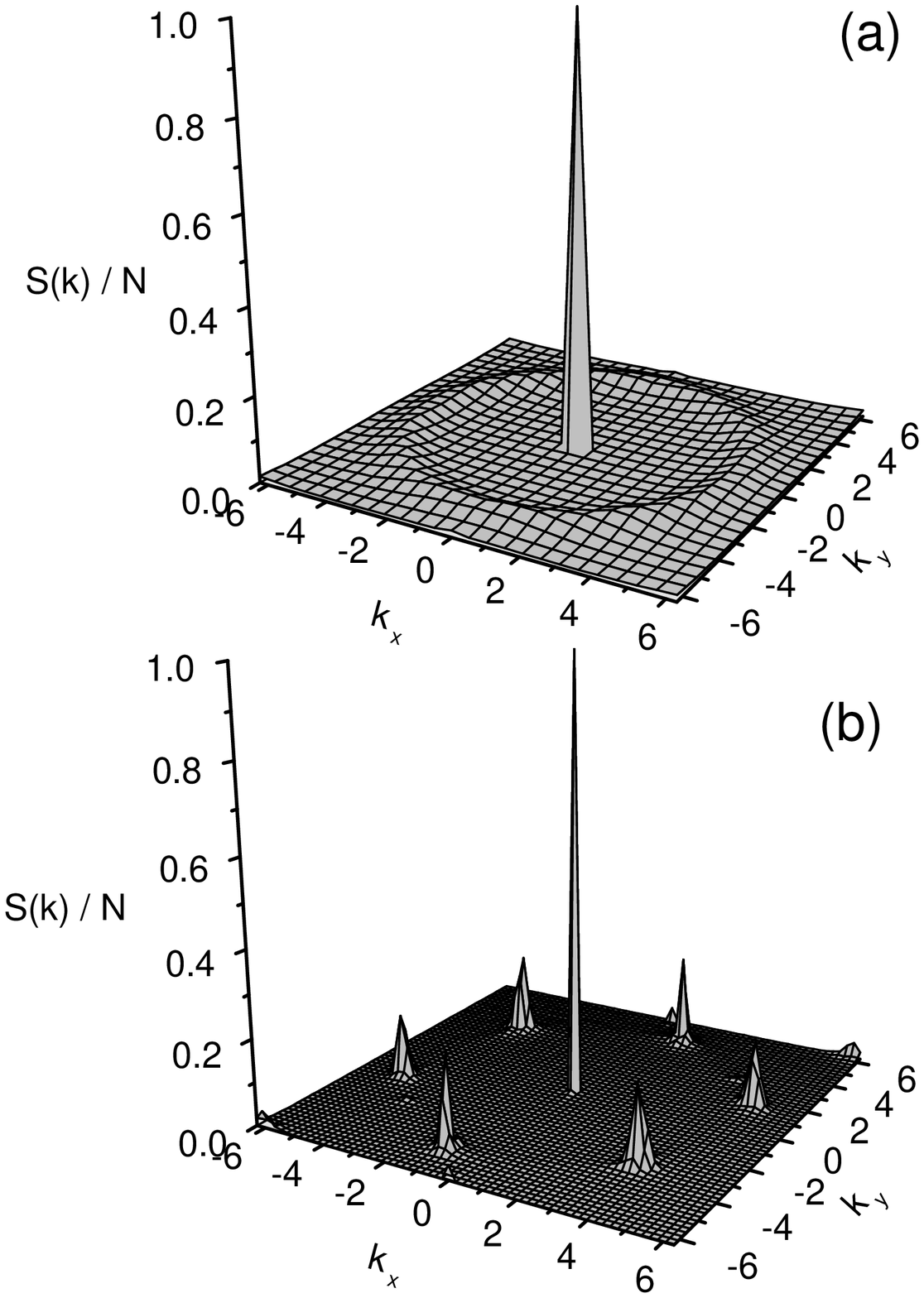}
\caption{ Structure factor $S(k)/N$ at $T=0.2$, $F_x=1.5 $,
$F_y=0$ for the periodic substrate and fixing one of the
adsorbates at the potential minima for a system size $L=12$ (a)
and $L=30$ (b).} \label{sfimp}
\end{figure}

Finally, we like to examine how the results presented here depends
on the details of the modelling of the disorder and defect pinning
centers. In this regard, it is interesting to compare the results
presented here  with a previous work by Persson \cite{persson}
where the disorder  was implemented by fixing one of the adsorbate
atoms. It was found that in this case, corresponding to
$x_d=0.0069$ (linear size $L=12$), the effective temperature $T^*$
increases with increasing force for $F> F_a$ and that the sliding
incommensurate solid phase observed for the periodic substrate is
destroyed and replaced by a fluid state.  At first sight this
seems to be in total contrast to the results found in this work
for the corresponding dilute disorder limit. To understand this
difference, we  note that in  a strict two-dimensional model, all
the adsorbate particles are confined to move in the $x$-$y$ plane
only. Hence, the structure of the sliding layer has to undergo
strong distortions in the vicinity of the fixed impurity atom, due
to the hard core repulsion of the Lennard-Jones interacting
potential, leading to a fluid phase. Thus, in this fixed-impurity
model, even a low concentration of impurities actually correspond
to strong disorder, in sharp contrast to the model studied here.
One would expect that even for the fixed impurity model,  a very
small concentration of impurities should lead to a more ordered
sliding state. To verify this, we have computed the structure
factor for the fixed impurity model at two concentration of
$x_d=0.0069$ ( linear size $L=12$) and $x_d=0.0012$ ( linear size
$L=30$) in the steady sliding state at $F_x=1.5$. For $L=12$ as
shown in Fig. \ref{sfimp}a, the corresponding structure factor is
characteristic of a fluid phase, in agreement with the findings of
Persson \cite{persson}. However, for the larger system size $L=30$
in Fig. \ref{sfimp}b, and therefore more dilute disorder, sharp
peaks develop corresponding to an hexagonal structure. This
structure is very similar to what we found in the present model
for $x_d=1$ as shown in Fig. \ref{sf}b indicating that the
overlayer in this situation corresponds to an incommensurate
hexagonal floating phase.

\subsection{Transverse response at large driving force}

In this section, we focus our attention on the nature of the
sliding state at large driving forces $F_x > F_c$, and its
relation to the response of the sliding layer to an additional
transverse driving force. It turns out that the nonlinear response
of the overlayer in the transverse direction is sensitive to the
nature of the sliding state, thus serving as an ideal probe for
the structure of the sliding state which is otherwise difficult to
study.

\begin{figure}
\includegraphics[bb=1cm 3cm 20cm 28cm, width=8cm]{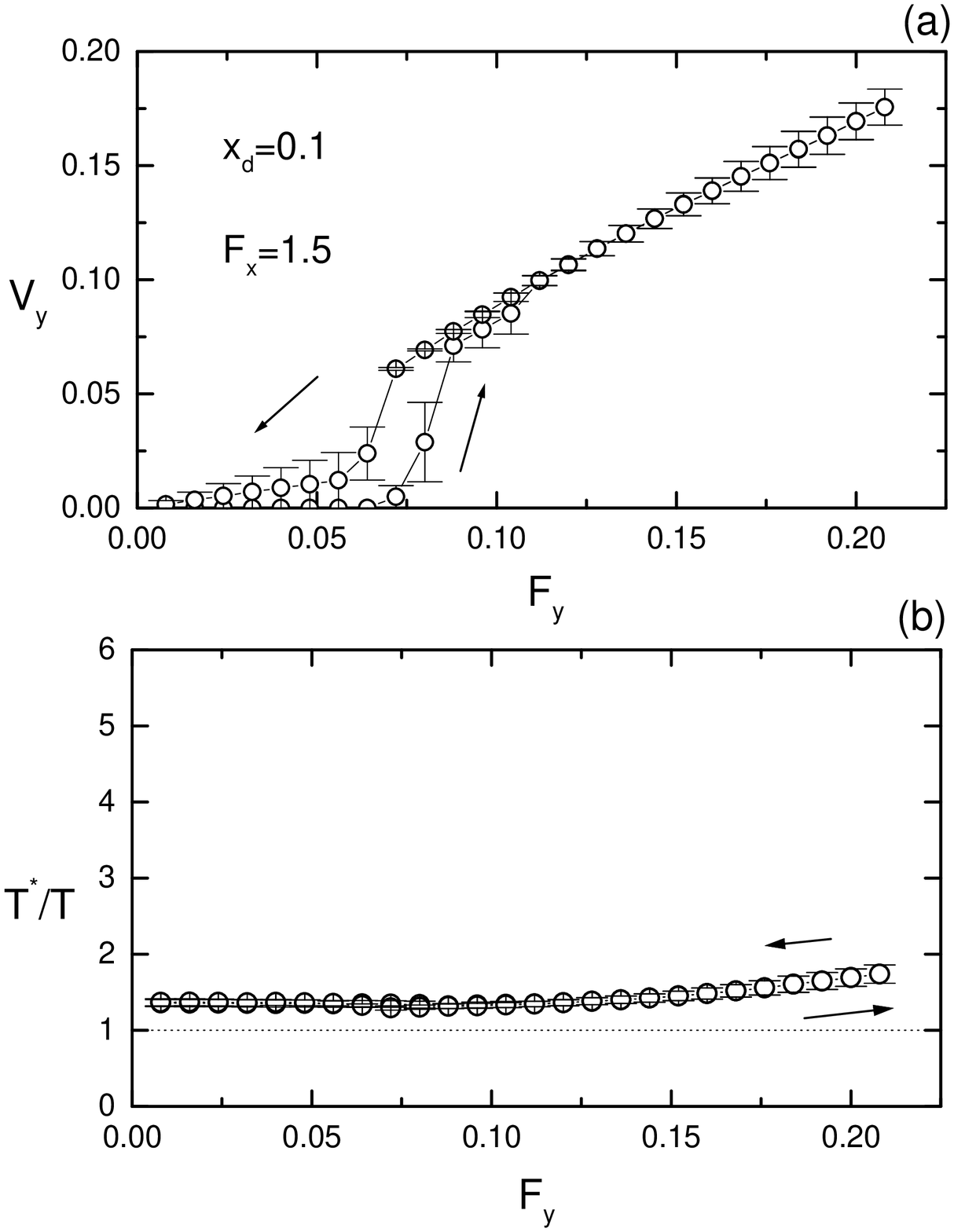}
\caption{ Transverse velocity $V_y$ (a) and effective temperature
$T^*$ (b) of the overlayer as a function of an additional force
along the $y$-axis at fixed $F_x$, for dilute disorder $x_d=0.1$.
Arrows indicate the direction of force variation. Results are for
$F_x=1.5$, $T=0.2$, and system size $L=20$. } \label{tx01}
\end{figure}

\begin{figure}
\includegraphics[bb=1cm 3cm 20cm 28cm, width=8cm]{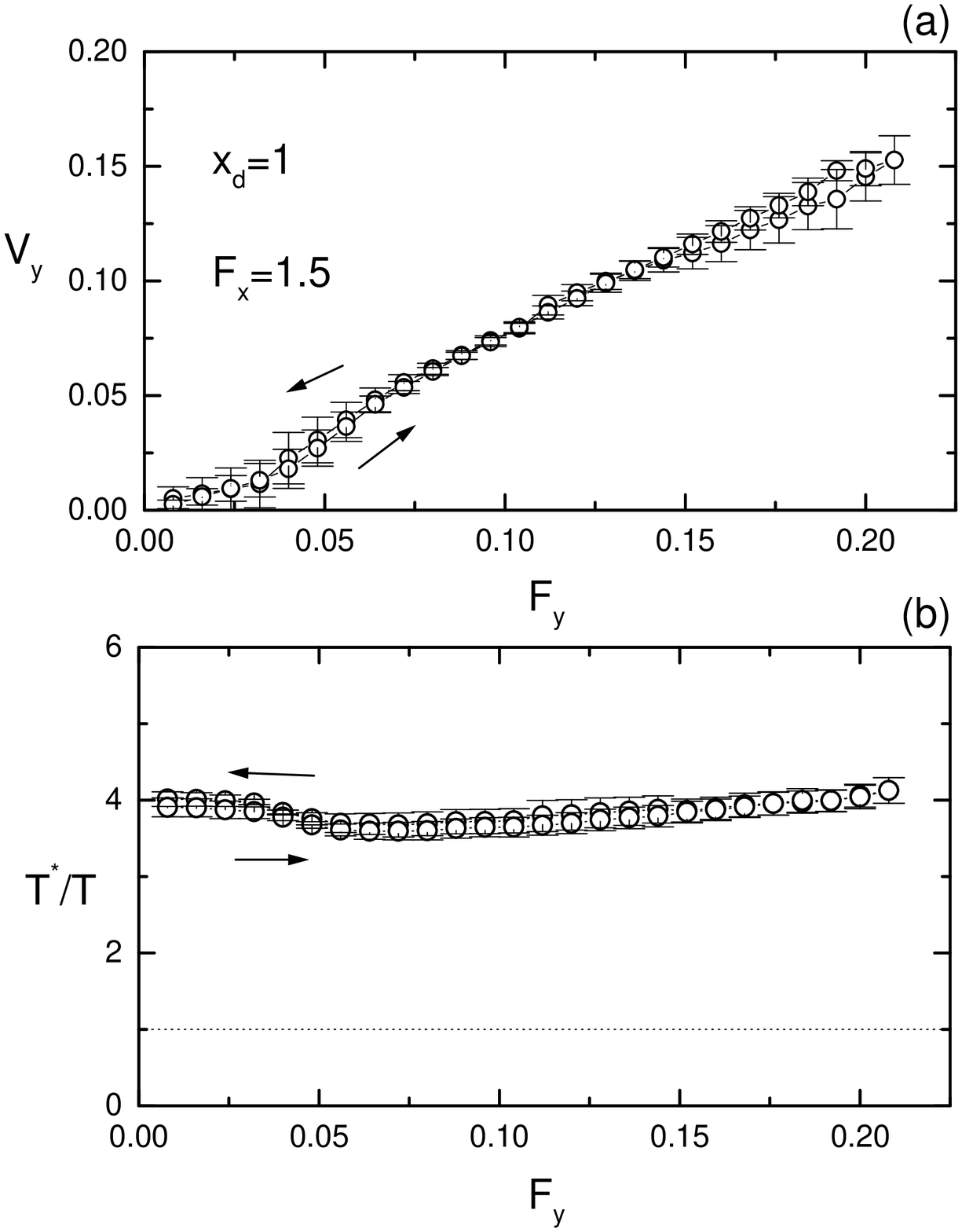}
\caption{Transverse velocity $V_y$ (a) and effective temperature
$T^*$ (b) of the overlayer as a function of an additional force
along the $y$-axis at fixed $F_x$, for dense disorder $x_d=1$.
Arrows indicate the direction of force variation. Results are for
$F_x=1.5$, $T=0.2$, and system size $L=20$. } \label{tx1}

\end{figure}

We have studied the transverse response of the overlayer in a
steady sliding state along the x direction. For a fixed large
driving force  $F_x=1.5 > F_c$, an additional variable transverse
force $F_y$ is applied, and the average transverse drift velocity
$V_y$ as a function of $F_y$ is determined. The resultant
transverse response is shown in Fig. \ref{tx01} and \ref{tx1} .
For dilute disorder $x_d=0.1$, we find that there is a transverse
depinning transition with a critical transverse force $F_{ya}$.
This transverse depinning transition  is accompanied by hysteresis
behavior with the lower critical force $F_{yb} \sim 0$. The
existence of this transverse depinning transition for weak
disorder is consistent with our finding of the nature of the
sliding state at large longitudinal driving force. As shown in
Fig. \ref{lx01}, for the sliding state with $F_y=0$ and $F_x >
F_c$, we have $T^* \sim T$. The structure factor shown for this
state in Fig. \ref{sf} has prominent peaks at the $c(2x2)$
commensurate structure despite the disorder resulting from a
random distribution of  the pinning centers. Thus, even while
sliding along the x direction, the overlayer can be pinned in the
transverse direction by the commensurate pinning potential. This
is origin for the existence of the transverse depinning
transition. In the other limit of dense pinning centers $x_d=1$
corresponding to strong disorder, the transverse response in Fig.
\ref{tx1} shows no hysteresis and the transverse critical force
$F_{ya}$ appears to vanish. Again, this behavior can be understood
from the nature of the corresponding sliding state at $F_y=0$. As
shown in Fig. \ref{lx1}, in this regime of strong disorder, $T^*
>> T$ when $F_x > F_c$.
Here the overlayer film is fully disordered and incommensurate
with the substrate along both the x and y directions. As a result,
there is no pinning force along the transverse direction, hence
the absence of a transverse depinning transition.

\section{Conclusions}

We have studied the nonlinear sliding friction  of adsorbed
overlayers on disordered substrates with point defects (pinning
centers). A simple  model is introduced which is a generalization
of the driven monolayer model previously used for periodic
surfaces \cite{book,persson,gy00}. The present model incorporates
in a convenient way the effects of point defects on an initially
ordered substrate. We have studied the longitudinal and transverse
response in this model by Brownian molecular dynamics simulations
in both the weak and strong disorder limit. The results show that
the main features, such as hysteresis, longitudinal and transverse
depinning previously observed for periodic surfaces
\cite{persson,gy00}, still survive in presence of weak disorder
but are washed out in the strong disorder limit. The results for
longitudinal response bear a strong qualitative resemblance to the
recent experimental data of the depinning transition of  Kr films
on Gd, using the quartz crystal microbalance technique
\cite{carlin}. This qualitative agreement supports the
interpretation \cite{carlin} of the experimental data that the
aging effects on the depinning transition are due to the
increasing disorder of the substrate. The present model can also
be used to gain insight into other interesting features of the
experimental results such as the coverage dependence of the
depinning transition and sliding friction. We expect that the
coverage dependence in the weak disorder limit to be similar to
that obtained in numerical simulations for a periodic substrate
\cite{persson} but quite different in the strong disorder limit.
Moreover, within the present model, a study of the nucleation of
topological defects and detailed microscopic mechanism of the
depinning transition itself should be intesting for driven lattice
systems in general.

\section{Acknowledgments}

This work was supported by a NSF-CNPq international collaboration
grant (SCY and E,G,) and  Funda\c c\~ao de Amparo \`a Pesquisa do
Estado de S\~ao Paulo - FAPESP (E.G.)(Grant no. 03/00541-0).

\end{document}